\documentclass[twocolumn,showkeys,preprintnumbers]{revtex4-2}
\usepackage{graphicx}
\usepackage{epstopdf}

\newcommand{\narrowimagesize}{5.8cm} 
\usepackage{soul}
\usepackage{color}
\begin{document}

\title{Quodon Current in Tungsten and Consequences for Tokamak Fusion Reactors}

\author{F. Michael Russell$^{1}$, Juan F. R. Archilla$^{2,*}$, Jos\'e L. Mas$^{3}$ }
\affiliation{$^{1}$School of Computing and Engineering, University of Huddersfield, HD1 3DH, United Kingdom\\
$^{2,*}$Group of Nonlinear Physics, Department of Applied Physics I, Universidad de Sevilla, ETSI Inform\'atica, Avda Reina Mercedes~s/n, 41012-Sevilla, Spain\\
$^{3}$Group of Applied Nuclear Physics, Department of Applied Physics I, Universidad de
Sevilla, ETSI Inform\'{a}tica, Avda Reina Mercedes s/n, 41012-Sevilla, Spain
}
\email[Corresponding author: ]{archilla@us.es}

\begin{abstract}
Tokamak fusion reactors produce energetic He ions that penetrate surfaces less than 20 micron and neutrons that spread throughout the reactor. Experiments with similar swift He ions in heavy metals show that the vibronic coupling of nonlinear lattice excitations creates mobile lattice excitations, called quodons. These are decoupled from phonons, move ballistically at near sonic speed and propagate easily in metals and insulators. They can couple to and transport electric charge, which allows their observation in experiments. They rapidly disperse heat throughout a fusion reactor and carry charge through electrical insulators.  In this paper we present an experimental design that separates quodon current and conduction current and therefore makes it possible to measure the former. We also present time-of-flight experiments that lead to an estimation of the quodon speed which is of the order of the sound velocity and therefore much faster than the drift of electrons or holes in conduction currents.  We present results on quodon current in tungsten, a material widely used in nuclear fusion technology, showing that many quodons will be produced in fusion reactors. It is predicted that at high output powers, quodons created by He ions and neutrons might adversely impact on cryogenic systems.
\end{abstract}


 \maketitle

\section{Introduction}
Fusion of deuterium and tritium produces 3.5 MeV He ions and 14 MeV neutrons. Fusion reactors extract as much as possible of this energy to generate electrical power and generate tritium from the lithium isotope $^6$Li by neutron capture. Neutrons disperse their energy throughout the reactor, making it difficult to isolate and utilize their energy, leaving He ions as the dominant source of energy for generation of electrical power. Both particles will inevitably hit and penetrate metallic surfaces where they lose energy by interactions with atoms as they propagate. It is the nature of these interactions that we examine here, especially nonlinear interactions associated with energetic particles.

Swift charged or uncharged particles can cause interstitials, ionization, plasmas and phonons when passing through a solid\,\cite{fleischer2003,lazarev-bakai2013}.
 All of these except phonons are highly localized to the path of the moving particle. Typically, a 3.5 MeV He ion penetrates less than 20 microns in a metal. In a fusion reactor this causes intense heating of the irradiated surfaces. It has been known for several decades that essentially loss-free transport of kinetic energy is possible by mobile entities called quodons, which are created by nonlinear atomic interactions involving energetic particles propagating in a solid\,\cite{russell-release1993,schlosser1994,russell-collins95a,russellnaturereview2022}. The essential condition for their creation is a highly localized interaction as opposed to distributed excitation. The collision of a swift particle with a lattice atom can create a mobile discrete-breather excitation\,\cite{flachmoving1999,aubry2006,yoshimura-doi2007,flach2008,kistanov2014,bajars-physicad2015,doi-yoshimura2016,bachurinagmoving2018}, also called an intrisic localized mode\,\cite{sievers-takeno1988,sato2015}.

In a phyllosilicate crystal with layered structure, a useful model for a quodon is a discrete breather with an inner structure of longitudinal optical-mode oscillations. The study of quodons began with the observation of dark linear defects in crystals of the phyllosilicate mineral muscovite mica. The finding that some of these lines showed evidence of unique Rutherford scattering showed that those lines were fossil tracks of intermediate-energy positrons emitted by decay of the potassium isotope $^{40}$K atoms in the crystals\,\cite{russell67a,russell88-identification,russell-archilla-chaos2021}. Later the fossil tracks of relativistic muons and electron-positron showers were identified\,\cite{russell67b}. However, most of the lines were of unknown origin. They lay in principal crystallographic directions with lengths limited only by size of the crystals, the longest found of 0.6 m length. All the lines and tracks are decorated with the mineral magnetite, which is precipitated as the crystals slowly cool at temperatures above 700 K\,\cite{russell68,russell-decorated91a}. Only moving positive charges trigger this precipitation\,\cite{russell88-positive,russell88-identification}. It followed that the unexplained lines were due to moving positive charges\,\cite{russell-tracks-quodons2015article,archillaLoM2016}. The existence of the long lines and the conditions of their creation showed that the moving causative entity, a quodon, was decoupled from phonons and was tolerant of minor lattice defects such as dislocations, vacancies and interstitials.

The reality of quodons, capable of propagating many orders of magnitude further than the stopping range of 3.5 MeV He ions in a solid, was demonstrated experimentally. An edge of a muscovite crystal held in a vacuum was bombarded at near grazing incidence with 5 MeV He ions. It caused atoms to be ejected from the opposite edge of the crystal 7\,mm distant from the irradiated edge\,\cite{russell-experiment2007}. To first order, the energy of a quodon causing ejection should be at least the sublimation energy of the ejected potassium atoms of 0.89 eV. Later was it shown that quodons can capture, and carry for considerable distances, a negative or positive charge at ambient temperatures without any applied voltage to drive the electric current\,\cite{russell-archilla2017,russell-archilla2019}.The existence of fossil quodon tracks in multiple principal crystallographic directions in absence of an applied emf is inconsistent with conduction currents. Charged quodons are the consequence of vibronic coupling of lattice nonlinear vibrations with an electron or hole\,
\cite{vekhter1995,flach1996,brizhik2003,henning-prb2006,hennig2007,chetverikov-entropy2016,velardeJChP2020}. For simplicity, quodons behave like nano-scale atomic tsunamis or shockwaves on which a charge can surf.

Quodons were largely unknown outside of nonlinear studies for several reasons. Firstly, they occur within a condensed material and when neutral, that is, carrying no charge, are undetectable. Only after they capture a charge or eject an atom by inelastic scattering can they be detected. Secondly, it was possible that conduction currents may contribute to currents attributed to charge carried by quodons. Only recently was this dilemma eliminated by development of the triple-filter technique (TFT) explained below. Thirdly, it was found that quodons can carry charge through insulators, thereby shorting quodon-currents to ground in coaxial cables\,\cite{russell-archilla2021rrl,russell-archilla2022ltp}. Also, there are few situations where significant fluences of quodons are created by high energy ions irradiating a surface. Examples are ion implantation systems, proton beams for tumour therapy and fusion reactors. Nonlinear atomic interactions also will occur in experiments involving relativistic particles but may be missed by the low fluences involved.

\section{Separation of Quodon from Conduction Currents}
\label{sec:triplefilter}
The possibility of conduction currents contributing to measured currents in irradiation experiments, perhaps also via surface currents, led to the development of the triple-filter technique (TFT). For the first time, this enabled quodons to be studied in any material. We first describe  the technique and then provide some experimental details and comments.

\subsection{The triple filter technique}

\begin{figure}[ht]
\begin{center}
  \includegraphics[width=\narrowimagesize]{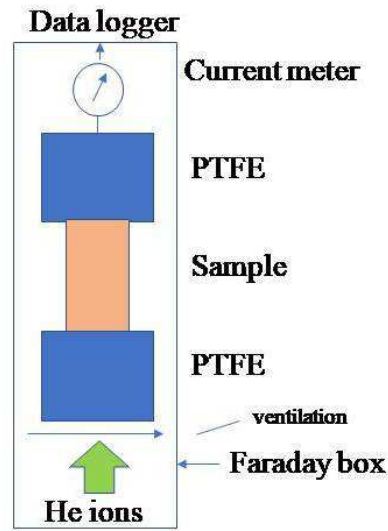}
  \caption{Diagram of the triple-filter-technique. The sample to be tested for existence of a quodon-current is sandwiched between two slabs of the electrical insulator polytetrafluorethylene PTFE. Both the volume and surface resistivity of PTFE are very high, requiring very large voltages to be applied to create even small conduction-currents. }
  \label{fig:triplefilter}
  \end{center}
  \end{figure}

The TFT is illustrated in Figure\,\ref{fig:triplefilter}. Swift particles, ions or atoms, bombard a slab of polytetrafluorethylene (PTFE). In absence of an applied emf, charge will accumulate at or near the bombarded surface, inducing a potential gradient across the slab, which is in contact with a sample to be tested for existence and propagation of a quodon-current. The opposite end of the sample is connected to a second slab of PTFE, the end surface of which is connected to a current meter. The meter will measure an induced current and any current carried by quodons. But little or no conduction current will flow through the PTFE slabs. To achieve the observed maximum current by conduction would require an applied voltage approaching mega-volts. A typical plot of current, for a sample of the heavy metal tungsten, starts with small initial transient currents followed quickly by rapid increase to a large positive current that is hundreds of times larger than the current injected by the bombardment. After reaching a maximum the current slowly decreases to a limiting current equal to the current injected by the swift particles. This behaviour is inconsistent with a conduction-current.

Using this technique, forty different materials were tested, ranging from Aerogel to tungsten. All showed quodons could exist and propagate in condensed matter. That quodons transport heat energy in absence of a thermal gradient does not violate thermodynamics. The main difference from normal heat conduction processes is that quodons transport heat ballistically.

\subsection{Experimental details}

The physical size of all the components is limited by that of the Faraday box which is essential when measuring small currents. The box is 12\,cm tall with cross-section of 6\,cm$\times$6\,cm. There is no basic limit to the size of the parts used in the TFT. Most are cut from a block of PTFE using a tungsten carbide saw disc. Typically, they are of 1 to 2\,cm blocks that are simple to handle. Samples are typically between 1 and 10\,cm$^3$ in volume. The tungsten sample used in Sec.\,\ref{sec:tungsten} was a cylinder of 1\,cm diameter and 1.5\,cm length.

It is essential to achieve good surface contacts between parts. The most used method is an interference fit of a rod in a hole drilled in the PTFE slab. If the sample to be studied cannot be machined then it can be sandwiched between two slab of PTFE and clamped with a nut and bolt. Clean surfaces are not essential  because quodons can propagate through multiple thin layers of different materials.

Experiments were done at room temperature, that is, around 21\,C. Some test at 80\,C did not provide significant change.

Main source of electrical noise is from cosmic rays passing through the samples and blocked by PTFE. This is of the below $\pm 5$\,fA, for currents that can reach $10^4$\,fA, bringing about a very good signal-noise ratio.

\section{Quodon Time of Flight}
\label{sec:timeofflight}
\begin{figure}[thb]
\begin{center}
\includegraphics[width=\columnwidth]{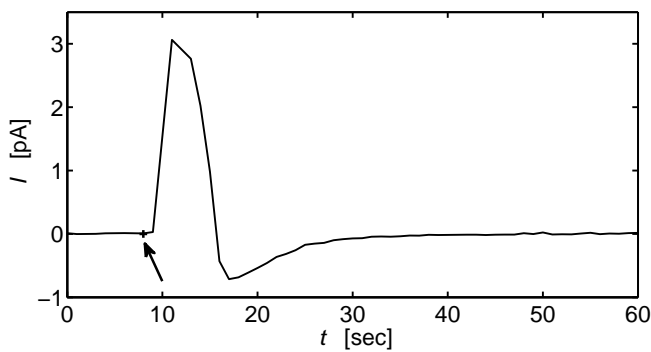}
\end{center}
\caption{Plot of current carried by quodons flowing through a sample of copper in a time-of-flight measurement when irradiated with 4 MeV He ions. The He ions embed first in a 2 cm slab of PTFE where they create quodons which then propagate through 1.1 km of copper and finally through a second 2 cm slab of PTFE. The quodon-current flowing through the sample was detected within 1 second of irradiation starting. The alpha-gate opened in less than 0.3 seconds.}
\label{fig_timeofflight}
\end{figure}

The approximate speed of quodons in various materials was examined by time-of-flight technique. Figure\,\ref{fig_timeofflight} shows the plot of quodon-current flowing through a copper wire following the start of irradiation with alpha particles at the time indicated by an arrow. The current was detected within 1 second of starting irradiation; it is limited by the shortest sampling rate of one second. Measurements on samples of iron and PTFE with much shorter flight paths gave the same shape of current plot, also limited by the sampling rate. The subsequent decrease in current is due to lack of free electrons available to be carried by the quodons not a decrease in the fluence of quodons, which stays constant.The speed of quodons in copper estimated in this way is of similar magnitude to the sonic speed of 2.3 km/s.

\section{Irradiation of Tungsten with 4 Mev He Ions}
\label{sec:tungsten}

\begin{figure}[thb]
\begin{center}
\includegraphics[width=\columnwidth]{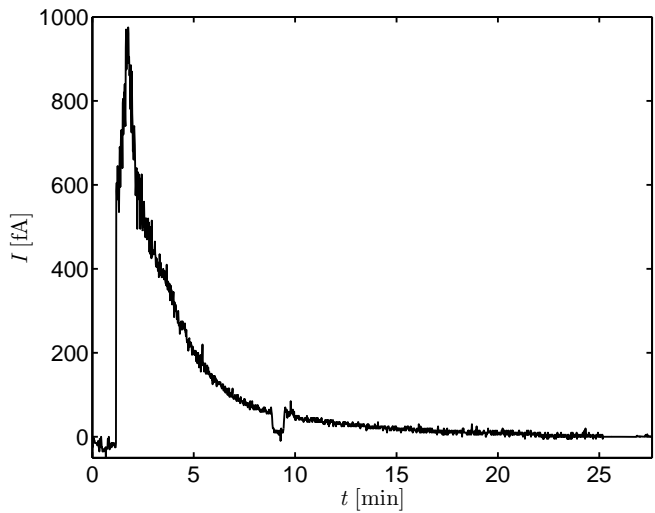}
\end{center}
\caption{Plot of quodon current when tungsten is irradiated with 2000 sec$^{-1}$ of 4 MeV He ions in the triple filter technique arrangement to eliminate conduction currents. The peak current is about 0.96\,pA. The number of quodons needed to transport this current is $6\times 10^8$\,s$^{-1}$.}
\label{fig_tungsten}
\end{figure}

To relate this work to the ITER prototype fusion reactor\,\cite{iterorg,claessensITER2020} in which He ions are directed to a diverter made of tungsten, the triple-filter-technique was used to measure the number of quodons created in a target of tungsten when irradiated with 4 MeV He ions. The current plot is shown in Figure\,\ref{fig_tungsten}. The peak current carried by quodons was 0.96\,pA requiring at least $5.9\times 10^6$ quodons per second. The
number of He ions hitting the target per second was about 2000, based on the activity of the $^{241}$Am source and the geometry of the source and target. The injected current is variable due to surface conditions and creation of secondary electrons.
The average number of quodons generated per He ion was about 3000. More might have been created but not observed due to limited availability of free charges, without which quodons cannot be detected. The ejection experiment, which involved sublimation energy of the ejected particle, showed that quodons could have an energy of at least 1\,eV after propagating 7\,mm in muscovite\,\cite{russell-experiment2007}. Studies of the creation of mobile optical-mode breathers showed that at sufficiently high energies of collision two quodons could be generated moving in opposite directions\,\cite{bajars2021}. The efficiency of energy-transfer from a swift particle to a quodon was found from numerical and analogue models to be about 50\%. This suggests quodons created by the maximum of 42 eV recoil energy released in beta-decay of $^{40}$K could have energies up to about 10 eV\,\cite{archilla-kosevich-springer2015article,archillaLoM2016}. For present purposes, we suppose that quodons can have energies in the range from 1 to 10\,eV. On this basis, the amount of energy of the He ions hitting the diverter that is transferred to quodons would be in the range 0.2\% to 2\%.

\section{Effect of Quodons on a Tokamak}
As quodons can transport both kinetic energy and charge there are two aspects to their influence in fusion reactors. Most quodons created by high energy particles are neutral but they can acquire an electric charge in the material they move through. We examine first the transport of kinetic energy. Based on experiments with 5\,MeV He ions, quodons can exist and propagate through 0.7\,m of PTFE, 2\,m of iron and 1.1\,km of copper wire\,\cite{russell-archilla2022ltp}. So far, no condensed material has been found in which quodons cannot exist. For the first time, we use these results to examine the possible consequences of the generation of quodons in prototype tokamak fusion reactors. For this purpose, we suppose that quodons propagate ballistically at near sonic speed for distances comparable to prototype reactor structures. Since the aim of the reactors is to generate electricity, we examine the limiting case where all the energy of the He ions is deposited in metals and then transferred to steam turbines. This ignores their energy spectrum from degradation in the plasma. We do this because, for given electrical power output, any reduction in the energy of the He ions leaving the plasma requires an increase in their numbers and thus of quodons.

Consider the ITER tokamak\,\cite{miyamoto2016,itertokamak2023}designed to produce 500\,MW of thermal heat.The difficulty of harnessing the energy of neutrons means most of the thermal energy available for generating electricity comes from the He ions. These contribute about 20\% of the heat energy or 100\,MW\,\cite{itertokamak2023}. If future designs of electrical generators achieve 45\% thermal efficiency, then ITER could generate 45\,MW of electrical power. On the basis of He ions delivering 100 MW to the diverter then the power transported by quodons is in the range 0.2 to 2\,MW. This will be spread swiftly through the nearby reactor components including the superconducting magnet system cooled to 4.5\,K, which has a refrigeration power of 75\,kW\,\cite{itercryo2023}. The ability of quodons to propagate through metals and thermal insulators means they can reach all parts of the magnet systems. We propose the ballistic transport of kinetic heat by quodons could compromise the performance of magnets.

This study of the creation of quodons and their propagation in condensed materials is based on experiments conducted with very small irradiation current fluences. Fortunately, quodons can trap a charge which allows their presence to be measured as a current. Since their creation depends on two-body interactions as opposed to distributed excitation, linear scaling to vastly greater ion fluences is expected. An upper limit might be approached when a quodon is created near another one in flight, which is highly unlikely in present designs of fusion reactors. Quodons also transport charge through electrical insulators, which might increase there impact on magnet systems.

No material has been found that does not allow quodons to propagate. This is expected, as quodons can be seen as small packets of energy, they are internally reflected at a vacuum interface and will pass easily through any insulating materials used in ITER. Therefore, at present, it seems difficult to shield parts of ITER from quodons.  In our opinion, further experimentation on quodon control is an urgent matter.

\section{Conclusion}
\label{sec:conclusion}
Experiments have shown that irradiating solid materials with helium ions of up to 5\,MeV energy creates mobile nonlinear lattice excitations, called quodons, that move ballistically. They can propagate appreciable distances, of similar scale to components in fusion reactors, in contiguous condensed materials at ambient temperatures. Propagating ballistically, they transport heat much faster than by thermal conduction. Development of the triple-filter-technique allowed the study of quodons in metals as well as through conventional thermal insulating materials. As the thermal power created in fusion devices increases towards that needed for commercial generation of electricity, the flux of quodons will also increase. As there is no known way to stop or exclude quodons from regions consisting of the contiguous condensed materials, large quodon fluences may lead to overloading of cryogenic systems. Quodons can also trap and carry electric charge through electrical materials ballistically at near sonic speed, much faster than electrons drift in conduction currents. They have the unique ability to transport charge through electrical insulators. It is not known if this will impact adversely on superconducting systems used in tokamaks nor if quodons  can disrupt Cooper pairs.

\section*{Acknowledgments}
FMR wishes to thank Chris Eilbeck, Leonor Cruzeiro and Larissa Brizhik for useful and enlightening discussions on both the dynamic and electronic aspects of quodon interactions with matter.

\section*{Funding}
JFRA wishes to thank projects MICINN PID2019-109175GB-C22 and Junta de Andaluc\'ia US-1380977 and a travel grant from VII-PPITUS 2023.

\bibliography{osaka2023RRL,russell2023}
\end{document}